\documentclass[amsmath,amssymb,showpacs,reprint,superscriptaddress,aps,prl]{revtex4-1}
\usepackage[dvips]{graphicx}
\usepackage{subfigure}
\usepackage{amsthm}
\usepackage{amsfonts}
\newcommand{\ii}[0]{\imath}
\newcommand{\eps}[0]{\varepsilon}

\newcommand{\bra}[1]{\left<{#1}\right|}
\newcommand{\ket}[1]{\left|{#1}\right>}

\newcommand{\beq}[0]{\begin{equation}}
\newcommand{\eeq}[0]{\end{equation}}
\newcommand{\beqa}[0]{\begin{align}}
\newcommand{\eeqa}[0]{\end{align}}

\newcommand{\HC}[0]{\text{H.C.}}

\begin{document}
\title{Dynamical control of quantum state transfer within hybrid open systems}

\author{B. M. Escher}
\affiliation{Instituto de F\'isica, Universidade Federal do Rio de Janeiro, 21.942-972, Rio de Janeiro (RJ) Brazil}
\affiliation{Department of Chemical Physics, Weizmann Institute of Science, Rehovot 76100, Israel}
\author{G. Bensky}
\author{J. Clausen}
\author{G. Kurizki}
\affiliation{Department of Chemical Physics, Weizmann Institute of Science, Rehovot 76100, Israel}
\author{L. Davidovich}
\affiliation{Instituto de F\'isica, Universidade Federal do Rio de Janeiro, 21.942-972, Rio de Janeiro (RJ) Brazil}

\date{\today}

\begin{abstract}
  We analyze quantum state-transfer optimization within hybrid open
  systems, from a ``noisy'' (write-in) qubit to its ``quiet''
  counterpart (storage qubit). Intriguing interplay is revealed
  between our ability to avoid bath-induced errors that profoundly
  depend on the bath-memory time and the limitations imposed by
  leakage out of the operational subspace.  Counterintuitively, under
  no circumstances is the fastest transfer optimal (for a given
  transfer energy).
\end{abstract}


\maketitle
%\doublespacing

Commonly, manipulations of quantum information can be schematically
divided into three stages: ``writing-in'', ``storage'', and
``reading-out'' \cite{Nielsenbook01}. Realistically, some systems are
better suited for writing-in or reading-out than for storage, and vice
versa. This has prompted the suggestion of {\em hybrid}, composite
quantum systems \cite{Petrosyan09,Imamoglu09,Rabl06,*Tordrup08,Verdu09}: quantum operations are rapidly
performed and efficiently written in a qubit susceptible to
decoherence, e.g., a Josephson (superconducting) qubit; then the
quantum information is transferred (directly or via a link) to a
storage qubit resilient to decoherence (encoded in an ensemble of,
e.g., ultracold atoms);
then, on demand, transferred back and read-out from the same fragile
qubit. Our aim is to examine a strategy for {\em maximizing} the
average fidelity of quantum state-transfer in such hybrids, from a
subspace fragile against decoherence to a robust subspace, by choosing
an appropriate dynamical control field.

%\section{State transfer from a fragile qubit to a robust qubit}

To this end, we resort to a novel general approach to the control of
{\em arbitrary} multidimensional quantum operations in open systems
described by the reduced density matrix $\hat\rho(t)$: if the desired
operation is disturbed by linear couplings to a bath, via operators
$\hat S\otimes\hat B$ (where $\hat S$ is the traceless system
operator, and $\hat B$ is the bath operator), one can choose controls
to maximize the operation fidelity according to the following recipe,
which holds to second order in the system-bath coupling (Suppl.
Info.): (i) The control (modulation) transforms the system-bath coupling
operators to the time-dependent form $\hat S(t)\otimes\hat B(t)$ in
the interaction picture, via the rotation matrix $\eps_i(t)$: a set of
time dependent coefficients in the operator basis $\hat\sigma_i$
(Pauli matrices in the case of a qubit), such that:
\begin{equation}
\hat S(t)=\sum_i\eps_i(t)\hat\sigma_i.
\end{equation}
(ii) This allows to write the time-independent {\em score matrix},
describing how the fidelity scores (changes) for each pair of basis
operators applied:
\begin{equation}
  \label{eq-Gamma}
  \Gamma_{ij}\equiv\overline{\bra\psi[\sigma_i,\sigma_j\ket\psi\bra\psi]\ket\psi},
\end{equation}
where the overline is an average over all possible initial states.
(iii) Using $\Gamma_{ij}$ one arrives at a simple expression for the
average fidelity of {\em any} desired operation (within the stipulated
second-order accuracy):
\begin{equation}
  \label{eq-FG}
  \begin{split}
    f_{\text{avr}}(t)=&1-\int_{-\infty}^\infty d\omega G(\omega)F(t,\omega),\\
    F(t,\omega)\equiv&t^{-1}\eps_{t,i}(\omega)\Gamma_{ij}\eps_{t,j}^*(\omega),
  \end{split}
\end{equation}
where $\eps_{t,i}(\omega)$ is the finite-time Fourier-transform of the
rotation matrix $\eps_i(t)$, and the coupling spectrum $G(\omega)$ is
the Fourier-transform of the bath memory (correlation) function $\left<\hat
Be^{\ii\hat H_B t}\hat Be^{-\ii\hat H_B t}\right>$. Namely, the modulation
(control) spectrum $F(t,\omega)$ is defined according to the
operation, via the $\Gamma_{ij}$ score matrix. 
(iv) This fidelity is maximized by the variational Euler-Lagrange
method \cite{Jens10,*Gordon08}, which minimizes the overlap between
$G(\omega)$ and $F(t,\omega)$ under the constraint of a given
control energy or action.

\begin{figure}[htp]
\centering
\includegraphics[width=\linewidth]{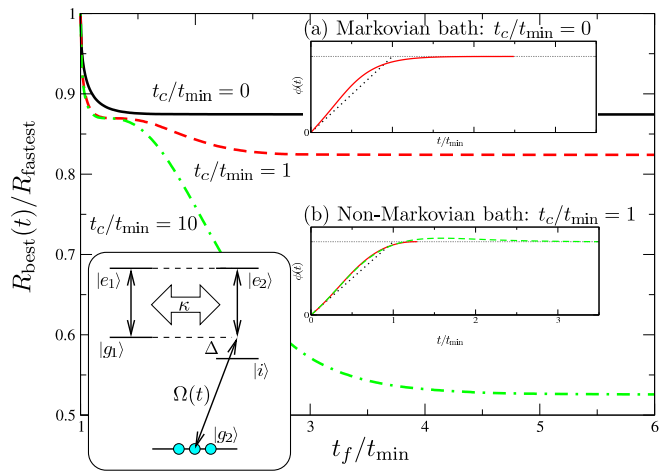}
\caption{Inset - scheme of coupling between ``noisy'' source qubit $1$
  and quiet target qubit $2$: $2$-photon transfer off-resonantly
  through $\ket i$ gives effective $\sigma_x^1\otimes\sigma_x^2$
  coupling with a controllable strength
  $V(t)=\frac{\kappa\Omega(t)}{\Delta}$, where $\Omega(t)$ is the Rabi
  frequency of an external laser field. Main panel: dependence of the
  lowest achievable average infidelity on the transfer time $t_f$
  normalized to the fastest transfer time $t_{\text{min}}$ at a given
  transfer energy (Eq.~\ref{eq-energycon}).  This function is plotted
  for various bath memory times: (black, solid) $t_c=0$
  (Markovian);(red, dash) $t_c=t_{\text{min}}$; (green, dash-dot)
  $t_c=10t_{\text{min}}$. Even for Markovian baths ($t_c\ll
  t_{\text{min}}$) the best solution is not the fastest one.  For
  non-Markovian baths ($t_c\gtrsim t_{\text{min}}$) two plateaux
  (regions of insensitivity to $t_f$) can be seen. The first plateau
  is independent of the memory time, and matches the Markovian
  plateau.  The second plateau is lower the longer the memory of the
  bath.  In (a) and (b) the transfer phase $\phi(t)$ is plotted versus
  $t/t_{\text{min}}$. The fastest modulation (black, dotted) with the
  Markovian optimal modulation (red, solid) and the non-Markovian
  optimal modulation (green, dashed), in Markovian (a) and
  non-Markovian (b) baths. In the Markovian bath the optimal
  modulation transfer starts off faster than the ``fastest'' transfer
  (when the information is still ``fresh''), and slows down
  subsequently. For the non-Markovian bath, optimal modulation
  achieves full transfer ($\phi(t)=\pi/2$) well within the modulation,
  but then ``overshoots'' ($\phi(t)>\pi/2$), and eventually returns to
  $\phi(t)=\pi/2$. }
\label{fig9}
\end{figure}

We use this general approach to optimize a reliable transfer of a
quantum state from a fragile qubit to a robust qubit. We choose to
focus on the case of two resonant qubits with temporally controlled
coupling strength. The free Hamiltonian without decoherence is then
\begin{equation}
  \label{eq-fullhamil}
  \begin{split}
    \hat H_S(t)=&\frac{\omega_0}{2} \left( \hat\sigma_z^{(1)} + \hat\sigma_z^{(2)} \right) + H_c(t),\\
    \hat H_c(t)=&V(t)\hat\sigma_x^{(1)}\otimes\hat\sigma_x^{(2)}
  \end{split}
\end{equation}
where $\hat H_c(t)$ is the Hamiltonian for the controlled interaction
between the qubits, $V(t)$ describing the ajustable amplitude of the
interaction (see Fig.~\ref{fig9}-inset, for an example where the
interaction amplitude is ajustable using an external laser field, as in \cite{Petrosyan09}). The
system-bath interaction Hamiltonian is taken to be
\begin{equation}
\hat H_I=\hat S\otimes\hat B(t)=\hat\sigma_{z}^{(1)}\otimes\hat B(t),
\label{eq-interaction}
\end{equation}
where $\hat B(t)$ is the bath operator $\hat B$ rotating with the free
bath Hamiltonian $\hat H_B$. This model represents proper dephasing in
the source qubit $1$ due to the bath operator $\hat B$, whereas the
target qubit $2$ is robust against decoherence. This model can be
generalized to {\em any degree of asymmetry} between the decoherence
properties of the two qubits.

Equations (\ref{eq-fullhamil})-(\ref{eq-interaction}) conserve the
parity of the number of excitations.
%while only the control $\hat
%H_c=\hat\sigma_x^{(1)}\otimes\hat\sigma_x^{(2)}$ can change the number
%of excitations. 
Hence, the full two-qubit system can be split into two subsystems
$\mathcal{O}=\text{span}\{\ket{g_1e_2},\ket{e_1g_2}\}$ and
$\mathcal{E}=\text{span}\{\ket{g_1g_2},\ket{e_1e_2}\}$, $\mathcal{O}$
and $\mathcal{E}$ standing for {\em odd} and {\em even} excitation
numbers, respectively:
\begin{equation}
\label{eq-Hamil-separated}
\begin{split}
\hat H_S+\hat H_I=&\hat H_\mathcal{O}+\hat H_\mathcal{E}\\
\hat H_{\mathcal{O}}=&V(t)\hat \sigma_x^{\mathcal{O}}+\hat \sigma_z^{\mathcal{O}}\otimes\hat B(t)\\
\hat H_\mathcal{E}=&\omega_0\hat\sigma_z^\mathcal{E}+V(t)\hat \sigma_x^{\mathcal{E}}+\hat \sigma_z^{\mathcal{E}}\otimes\hat B(t)
\end{split}
\end{equation}
where the appropriate Pauli matrices in the $\mathcal{O}$
($\mathcal{E}$) subsystems are:
$\hat\sigma_x^{\mathcal{O}}=\ket{g_1e_2}\bra{e_1g_2}+\HC$,
$\hat\sigma_z^{\mathcal{O}}=\ket{e_1g_2}\bra{e_1g_2}-\ket{g_1e_2}\bra{g_1e_2}$,
$\hat\sigma_x^{\mathcal{E}}=\ket{g_1g_2}\bra{e_1e_2}+\HC$ and
$\hat\sigma_z^{\mathcal{E}}=\ket{e_1e_2}\bra{e_1e_2}-\ket{g_1g_2}\bra{g_1g_2}$.
In essence we have one resonant and one non-resonant two-level
system, both coupled to the same dephasing bath, which renders them
inseparable. Both are subject to the same $\hat\sigma_x$ control, which
must be chosen to maximize the fidelity of a rotation in the
$\mathcal{O}$ subsystem while keeping the $\mathcal{E}$ subsystem
unchanged.

The accumulated phase
\begin{equation}
\phi(t) = \int_{0}^{t} V(t') dt'
\end{equation}
is our control function. In the ideal case, without decoherence or
leakage, the state transfer from qubit $1$ to qubit $2$ can be
perfectly realized if at the final time, $t_f$, the phase $\phi(t)$
satisfies $\phi(t_f) = \frac{\pi}{2}$, whence any initial state of
qubit $1$ is mapped onto that of qubit $2$ (initially in the ground
state)
\begin{equation}
\left( \alpha \vert g_{1} \rangle + \beta \vert e_{1} \rangle\right) \vert g_{2} \rangle \rightarrow \vert g_{1} \rangle \left( \alpha \vert g_{2} \rangle -i \beta \vert e_{2} \rangle\right) ,
\end{equation}
for any normalized $\alpha$, $\beta$. Here the states $\vert g_{1}
\rangle$ ($\vert g_{2} \rangle$) and $\vert e_{1} \rangle$ ($\vert
e_{2} \rangle$) are respectively the ground and the excited states of
the source qubit $1$ and the target qubit $2$.

There are two conflicting noise (error) considerations for the
transfer, each affecting a different subsystem: (i) In the presence of
interaction between the source qubit $1$ and the bath, the longer the
information stays in qubit $1$ the lower the fidelity of the transfer
(manifest in subsystem $\mathcal{O}$). (ii) On the other hand, if we
make the transfer extremely fast, it may result in population from
$\ket{g_1}\ket{g_2}$ leaking into $\ket{e_1}\ket{e_2}$, thus lowering
the fidelity of transfer (manifested in subsystem $\mathcal{E}$). Such
leakage\cite{Jens10,Wu09,*Wu02,*Byrd04,*Byrd05,*Lidar08} signifies the
violation of the rotating wave approximation (RWA). Namely, fast
modulation $V(t)$ may incur unwanted, off-resonant transitions if the
transfer rate is comparable to the energy difference (level distance)
of the qubits, $\omega_0$.

We first focus on bath-related errors (i), assuming that the RWA is
valid, i.e., there is no leakage because of the RWA violation. This
may be the case if the transfer time is much slower than the energy
separation $\omega_0$. This is also true when the non-RWA terms simply
do not exist, such as in $2D$ or $3D$ Heisenberg interactions (of the
form
$\hat\sigma_x^{(1)}\otimes\hat\sigma_x^{(2)}+\hat\sigma_y^{(1)}\otimes\hat\sigma_y^{(2)}$
or
$\hat\sigma_x^{(1)}\otimes\hat\sigma_x^{(2)}+\hat\sigma_y^{(1)}\otimes\hat\sigma_y^{(2)}+\hat\sigma_z^{(1)}\otimes\hat\sigma_z^{(2)}$,
respectively) where only number-conserving terms exist. The control
Hamiltonian $H_C(t)$ then has the RWA form \cite{Kofman04,*Gordon07}:
\begin{equation}
\label{eq-Hamil}
\begin{split}
\hat H_c(t)\equiv&V(t)\left(\ket{e_1g_2}\bra{g_1e_2}+\ket{g_1e_2}\bra{e_1g_2}\right)
\end{split}
\end{equation}
% In the frame rotating with $\omega_0$ the Hamiltonian (\ref{eq-Hamil}) becomes:
% \begin{equation}
%   \begin{split}
%     \hat H(t)=&V(t)\hat H_c+\hat S\otimes\hat B(t),\\
%     B(t)=&e^{\ii H_B(t)}Be^{-\ii H_B(t)}
%   \end{split}
% \end{equation}

The general expression (\ref{eq-FG}) derived from the score matrix
(\ref{eq-Gamma}) for the average fidelity of the transfer, completed
at $t_f$, is then (see Suppl.):
\begin{equation}
\label{eq-fidelity}
\overline{f(t_{f})} = 1 - \int_{-\infty}^{\infty} d \omega G(\omega) F(t_f,\omega),
\end{equation}
\begin{equation}
  \begin{split}
    F(t,\omega)=&\frac{2}{3}\left|\int_{0}^{t}d\tau\cos^2(\phi(\tau))^{2}e^{-i\omega\tau}\right|^2\\
    &+ \frac{1}{2}\left|\int_{0}^{t}d\tau \sin(2\phi(\tau))e^{-i\omega\tau}\right|^2,
  \end{split}
\end{equation}
being the transfer-oriented modulation control spectrum.

Thus, the average fidelity of the transfer has an involved dependence
on the modulation $V(t)$ and the transfer time $t_{f}$. The problem at
hand is to find the {\em optimal} transfer that minimizes the average
infidelity at time $t_f$, $1-\overline{F(t_f)}$.
% \begin{equation}
% 1-f_{av}(t_{f})=\int_{-\infty}^{\infty}d\omega G(\omega)F(t_f,\omega).
% \end{equation}
Obviously, zero infidelity is obtainable for infinitely fast
(zero-time) transfer, if we allow infinitely strong control.  Since
this is unphysical, we add a constraint on the total energy $E$
of the transfer process
\begin{equation}
\int_{0}^{t_{f}} dt \left(V(t)\right)^{2} = \int_{0}^{t_{f}} dt \left(\frac{d \phi(t)}{d t} \right)^2 = E.
\label{eq-energycon}
\end{equation}
As discussed below, this constraint can prevent leakage to levels out
of the operational qubit
subspace\cite{Jens10,Wu09,*Wu02,*Byrd04,*Byrd05,*Lidar08}. The
constraint defines the minimum possible time for the transfer
$t_{\text{min}}=\frac{\pi^2}{4E}$.

% In what follows we illustrate these general expression for two typical
% non-Markovian baths with finite memory times of $t_c$: (i) a bath with
% Lorentzian spectrum, i.e. an exponentially correlation function
% \begin{equation}
% \Phi(t) = \frac{\gamma}{t_{c}} e^{- \vert t \vert / t_{c}},
% \end{equation}
% (ii) an ohmic bath, whose spectrum has an upper exponential cutoff (XXX ???)
% \begin{equation}
% \Phi(t) = \gamma t_{c} \left( \frac{1}{t_{c} + i t} \right)^{2}.
% \end{equation}

We illustrate the general expressions
(\ref{eq-fidelity})-(\ref{eq-energycon}) for a typical non-Markovian
Lorentzian bath spectrum, i.e. an exponentially decaying correlation
function $\Phi(t) = \frac{\gamma}{t_{c}} e^{- \vert t \vert / t_{c}}$,
$t_c$ being the correlation (memory) time. One might expect that for
such a simple bath the best strategy is the fastest possible transfer
under the energy constraint, i.e. when the modulation is given by
$V(0\leq t \leq t_{\text{min}})= 2 E / \pi$.  Surprisingly, a {\em
  slower} transfer ($t_f>t_{\text{min}}$) with an appropriate
modulation $\phi(t)$ (detailed below) can improve the average fidelity
even for a {\em purely Markovian} bath, with negligible correlation
(memory) time $t_c/t_{\text{min}}\to0$, and more so for baths with memory times
longer than the transfer time, $t_c\gtrsim t_{\text{min}}$.
% Similar results are obtained for an Ohmic bath (Suppl. XXX).

When the bath is memoryless, i.e.  Markovian, this improvement is
limited, as shown in Fig.~\ref{fig9}, to about $12\%$. By comparing
the ``best'' solution to the ``fastest'' one (Fig.~\ref{fig9} (a),
(b)), one can see the the ``best'' solution starts off faster and then
slows down, being overtaken by the ``fastest'' solution only at
$t\approx0.9t_{\text{min}}$. This illustrates the source of the
Markovian noise reduction: the ``best'' solution starts off faster, so
as to transfer more of the information while it is still nearly
untainted by the bath. Obviously, towards the end it must slow down so
as to comply with the energy constraint, thus resulting in total
transfer time $t_f$ that is longer than the fastest time
$t_{\text{min}}$ for the given energy.

However, when the memory-time $t_c$ of the bath is comparable to or
larger than the characteristic transfer time $t_c\gtrsim
t_{\text{min}}$, a much larger improvement can be achieved (see
Fig.~\ref{fig9}). Remarkably, the best solution actually performs a
full transfer, $\phi(t)=\pi/2$, well within the modulation time, but
rather than stop at $\phi=\pi/2$ it then ``{\em overshoots}'' the
transfer, so that $\phi(t)>\pi/2$, and then returns slowly to $\pi/2$.
This can explain the source of the noise reduction --- when
``overshooting'', the information partially returns from the target
(storage) qubit to the source (noisy) qubit, but with a negative sign.
Hence, similarly to the ``echo'' method, the noise now operates in the
reverse direction, correcting itself, i.e., the non-Markov bath effect
is undone.  This requires transfer times {\em significantly} larger
than the minimal transfer time $t_{\text{min}}$, ranging from
$3t_{\text{min}}$ to even $10t_{\text{min}}$ or more, yet the fidelity
increases substantially (up to $50\%$ in Fig.~\ref{fig9}).

% \begin{figure}[htp]
% \centering
% \includegraphics[width=0.7\textwidth]{guy/best_solutions.eps}
% \caption{The best solution of $\phi(t)$ for the minimum possible time
%   (black), the first plateaux (red) and the second plateaux (green,
%   where applicable) for a Markovian bath (top) and a non-Markovian
%   bath (bottom). One can see how in the solution of the first plateaux
%   transfer begins faster than the ``fastest'' transfer (when the
%   information is still ``fresh''), and slows down later. On the second
%   plateaux there is a full transfer ($\phi(t)=\pi/2$) at the middle of
%   the solution, but then the solutions ``overshoots''
%   ($\phi(t)>\pi/2$), and returns later to $\phi(t)=\pi/2$. This
%   ``overshooting'' uses the memory of the bath to ``undo'' some of the
%   decoherence, resulting is much lower decoherence.}
% \label{fig10}
% \end{figure}

Using the Euler-Lagrange variational method one can find an analytical
solution for the optimal modulation phase $\phi(t)$, given a Markovian
bath at long times (see Suppl. Info.). This yields:
\begin{equation}
  \label{eq-optimalsolution}
  \begin{split}
    \frac{d\phi_M(x)}{dx}=&\sqrt{\frac{\sin^2(2\phi_M(x))}{2}+2\frac{\cos^4(\phi_M(x))}{3}},
  \end{split}
\end{equation}
with $\phi_M(0)=0$. Eq.~(\ref{eq-optimalsolution}) determines the
shape of $\phi_M(x)$ and its formal ``energy''
$e_M=\int_0^\infty|\phi_M'(x)|^2dx=1.038\dots$ (where both $x$ and
$e_M$ are dimensionless). The general Markovian optimal modulation at
infinite time for {\em any} energy $E$ is then
$\phi(t)=\phi_M(\frac{E}{e_M}t)$, with an infidelity of
$\gamma\frac{e_M^2}{E}=\gamma\frac{1.077\dots}{E}$, $\gamma$ being the
dephasing rate of the source qubit $1$. The fastest modulation with
energy $E$ has an infidelity of
$\gamma\frac{\pi^2}{8E}=\gamma\frac{1.233\dots}{E}$. This means that
the optimal modulation has about $12\%$ less infidelity than the
fastest modulation for the same energy.

Let us now take into account the breakdown of the RWA as a noise
source. A realistic coupling of the form
$\sigma_x^{(1)}\otimes\sigma_x^{(1)}$, as in (\ref{eq-fullhamil}), has
non-RWA terms of the form $\ket{g_1g_2}\bra{e_1e_2}$ which do not
conserve excitation. In all control scenarios such terms are either
discarded or, at best, any transfer of population via non-RWA terms is
considered alongside all other forms of leakage.
%However these terms are different than leakage to higher levels.
%For one, the non-RWA transfer keeps the excitation in the system's
%computational subspace, and hence the information is still available
%for manipulation (and possibly correction). In addition 
However, there is a drastic difference in timescale between the
non-RWA terms in (\ref{eq-fullhamil}) and leakage to higher levels:
the qubit level separation $\omega_0$ is often orders of magnitude
smaller than the separation to the next (non-qubit) level. Hence,
leakage to higher levels requires timescales that are orders of
magnitude smaller than those breaking the RWA.
%Solving the non-RWA timescales could then vastly increase the transfer
%speed and, with it, the fidelity of the transfer.

\begin{figure}[htbp]
  \centering
  \includegraphics[width=\linewidth]{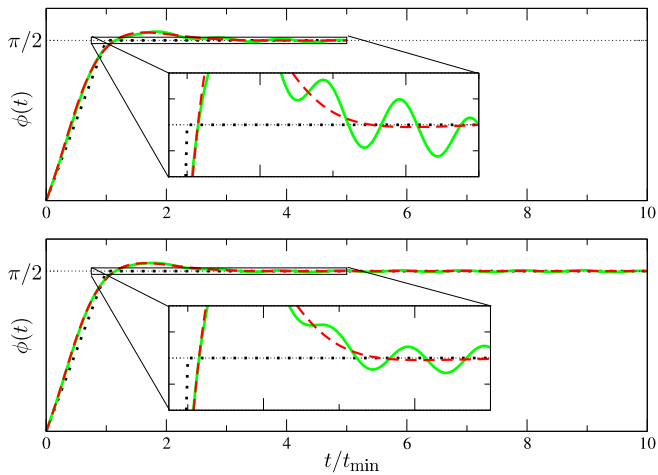}
  \caption{Fastest (black, dots) vs. best modulation for RWA (red,
    dash) and non-RWA (green) transfer with a final time of
    $t_f=5t_{\text{min}}$ (top) and $t_f=10t_{\text{min}}$ (bottom) in
    a non-Markovian bath. The level separation $\omega_0$ satisfies
    $\omega_0t_{\text{min}}=\pi$, giving $\sim2.5\%$ population
    leakage for fastest modulation and $\sim10\%$ loss of fidelity
    from decoherence. The sinusoidal ``wiggling'' in the non-RWA
    solution is larger when the final time is shorter (top).}
  \label{fig-nonRWQ}
\end{figure}

The RWA in (\ref{eq-fullhamil}) breaks down if the transfer time is
similar to or less than the inverse energy separation of the qubit,
$t_{\text{min}} \lesssim \omega_{0}^{-1}$.  This is a common case for qubits
whose resonance frequency is in the microwave (GHz) or RF (MHz) range.
In such cases, the optimization process must take into account both
the dephasing due to the bath as in
(\ref{eq-fidelity})-(\ref{eq-energycon}) and the error due to the
non-RWA terms when minimizing the infidelity.
% An energy constraint $E$ is still used to avoid leakage to
% higher (non-computational) levels.

If one of these errors is vastly larger than the other it is the only
one to consider. The problem changes only if the bath-induced and the non-RWA
errors are similar.
%
%The doubly excited state $\ket{e_1e_2}$ can only be populated using
%the non-RWA operator $\hat\sigma_x^\mathcal{E}$ in
%(\ref{eq-Hamil-separated}). 
In this case we find that dephasing of the doubly excited state caused
by $\hat\sigma_x^\mathcal{E}\otimes\hat B$ in
(\ref{eq-Hamil-separated}) is a fourth order effect and hence can be
ignored in the present second order treatment. The result of this
approximation is that the system can be split into two completely
separate subsystems $\mathcal{O}$ and $\mathcal{E}$, the former suffering
only from dephasing and the latter only from unwanted population of the
doubly excited level $\ket{e_1e_2}$. The Hamiltonians of these systems are
\begin{equation}
\begin{split}
\hat H_{\mathcal{O}}=&V(t)\hat \sigma_x^{\mathcal{O}}+\hat \sigma_z^{\mathcal{O}}\otimes\hat B,\\
\hat H_\mathcal{E}=&\omega_0\hat\sigma_z^\mathcal{E}+V(t)\hat \sigma_x^{\mathcal{E}}.
\end{split}
\end{equation}

The goal of our optimization is to find a control $V(t)$, shared by
both subsystems, which maximizes the fidelity of the transfer in
subsystem $\mathcal{O}$ (as per
(\ref{eq-fidelity})-(\ref{eq-energycon})) while at the same time
minimizing the doubly-excitation in subsystem $\mathcal{E}$.

The optimal modulation for Markovian and non-Markovian baths, for
different final times $t_f$, is given in Fig.~\ref{fig-nonRWQ}. The
result shows that the optimal modulation resembles the solution for
dephasing in Fig.~\ref{fig9}, but with added ``wiggles''. The
``wiggling'' takes up the entire time allowed for the transfer, but,
given more time for the total transfer, the amplitude of the
``wiggle'' diminish. This can be understood as follows: first you
should complete the transfer assuming the RWA so as to minimize the
information lost to the bath. Once the transfer is complete, and
decoherence is minimized, we can use whatever energy is left to return
the ``leaked'' excitation from $\ket{e_1e_2}$ back to $\ket{g_1g_2}$.
This can be done by a weak sinusoidal modulation of frequency
$2\omega_0$, inducing a Rabi coupling between the doubly excited and
zero-excited levels of $\mathcal{E}$. In practice this is not a
two-stage modulation, as the weak oscillation is superimposed on top
of the transfer modulation.

The energy needed to ``undo'' the non-RWA effect is inversely
proportional to the allowed time ---
$E\approx\frac{|\psi_{ee}|^2}{2t_f}$ (where $\psi_{ee}$ is the
amplitude of the doubly excited state $\ket{e_1e_2}$). 
%The longer the
%allowed total time $t_f$, the weaker the amplitude (and hence energy)
%of this sinusoidal modulation can be --- so that the total energy
%needed to undo the non-RWA effect is inversely proportional to the
%allowed time:
%\begin{equation}
%\begin{split}
%H=&\omega_0\hat\sigma_z+V\sin(2\omega_0t)\hat\sigma_x\\
%\Omega=&V;\quad T\approx\psi_{ee}V^{-1}\\
%E=&\int_0^TV^2\sin^2(\omega_at)dt\approx V^2\frac{T}{2}=\frac{|\psi_{ee}|^2}{2T},
%\end{split}
%\end{equation}
%where $\psi_{ee}$ is the amplitude of the doubly excited state
%$\ket{e_1e_2}$.
%
Hence, given enough time, the correction of the non-RWA effects
requires negligible energy, yielding the same results as in the RWA
case. If, however, time is limited --- a larger proportion of the
energy of the transfer must be reserved for the correction of the
non-RWA effect, resulting in smaller reduction of the bath-induced
noise.

To conclude, our analysis of state-transfer optimization within hybrid
open systems, from a ``noisy'' qubit to its ``quiet'' counterpart, has
revealed an intriguing interplay between our ability to avoid both
bath-induced errors that profoundly depend on the bath-memory time and
the limitations imposed by leakage out of the operational subspace.
Counterintuitively, under no circumstances is the fastest transfer
optimal (for a given transfer energy). Generalizations to
higher-dimensional cases are expected to follow analogous trends.

{\em Acknowledgments}: We acknowledge the support of EC (FET Open,
MIDAS project), ISF, DIP and the Humboldt-Meitner Research Award
(G.K.) and CNPq, CAPES, FAPERJ and the Brazilian National Institute
for Science and Technology on Quantum Information (B.M.E)

%\clearpage

\end{document}